\documentclass[
    ,final            
  ]
  {aipproc}

\layoutstyle{8x11single}

\usepackage{amssymb,amsmath,amstext,amsthm,amsfonts}

\def\be{\begin{equation}}
\def\ee{\end{equation}}
\def\bea{\begin{eqnarray}}
\def\eea{\end{eqnarray}}
\def\bear{\begin{array}}
\def\ear{\end{array}}
\def\bfig{\begin{figure}}
\def\efig{\end{figure}}
\def\bcen{\begin{center}}
\def\ecen{\end{center}}
\def\raw{\rightarrow}

\def\vr{\mathbf{r}}
\def\vp{\mathbf{p}}
\def\vk{\mathbf{k}}
\def\vkp{\mathbf{k'}}

\def\vq{\mathbf{q}}

\def\btau{\pmb\tau}
\def\bsigma{\pmb\sigma}

\def\bi{\begin{itemize}}
\def\ei{\end{itemize}}

\begin{document}

\title{Quasielastic Scattering at MiniBooNE Energies}


\author{L. Alvarez-Ruso}{
	address={Departamento de F\'isica, Universidad de Murcia, Spain}, 
  altaddress={Departamento de F\'isica, Centro de F\'isica Computacional, Universidade de Coimbra, Portugal}
}

\author{O. Buss}{
  address={Institut f\"ur Theoretische Physik, Universit\"at Giessen, Germany}
}

\author{T. Leitner}{
  address={Institut f\"ur Theoretische Physik, Universit\"at Giessen, Germany}
}

\author{U. Mosel}{
  address={Institut f\"ur Theoretische Physik, Universit\"at Giessen, Germany}
}

\begin{abstract}
We present our description of neutrino induced charged current quasielastic scattering  (CCQE) in nuclei at energies relevant for the MiniBooNE experiment. In our framework, the nucleons, with initial momentum distributions according to the Local Fermi Gas model, move in a density- and momentum-dependent mean field potential. The broadening of the outgoing nucleons due to nucleon-nucleon interactions is taken into account by spectral functions. Long range (RPA) correlations renormalizing the electroweak strength in the medium are also incorporated. The background from resonance excitation events that do not lead to pions in the final state is also predicted by propagating the outgoing hadrons with the Giessen semiclassical BUU model in coupled channels (GiBUU). We achieve a good description of the shape of the CCQE $Q^2$ distribution extracted from data by MiniBooNE, thanks to the inclusion of RPA correlations, but underestimate the integrated cross section when the standard value of $M_A = 1$~GeV is used. Possible reasons for this mismatch are discussed.
\end{abstract}

\maketitle


\section{Introduction}

By charged current quasielastic scattering (CCQE) one usually understands the reaction in which the elementary process
\be
\nu_l (p) \,+ \,n(k) \raw  l^-(k')\, + \, p(p') 
\label{qe}
\ee
takes place inside the nucleus. CCQE,  the largest reaction channel for $E_\nu \lesssim  2$~GeV, is of cardinal importance for oscillation experiments that rely on the detection of muons (electrons) in $\nu_\mu$ disappearance ($\nu_e$ appearance) searches. It is also the channel that can be more reliably used for a kinematical neutrino energy reconstruction, an indispensable exercise for a precise determination of oscillation parameters in long-baseline accelerator experiments. Moreover, quasielastic scattering is interesting by itself and has, indeed, been carefully investigated with electron beams both experimentally and theoretically with a large variety of models~\cite{Paviabook,Benhar:2006wy,Gil:1997bm,Amaro:2006if}. With neutrinos it is possible to study different properties of the nuclear response in the axial sector that are not (easily) accessible in electron scattering experiments. Provided that nuclear effects are under control, CCQE could be a source of information about the nucleon axial form factor, often parametrized as 
\begin{equation}
\label{FA}
F_A (Q^2) = g_A \left( 1 + \frac{Q^2}{M_A^2} \right)^{-2} \,,
\end{equation} 
where $Q^2 = -(k-k')^2$ and $M_A$ is the so called axial mass.

The definition of CCQE given above already implies the assumption that the neutrino-nucleus interaction takes place predominantly on a single nucleon (impulse approximation) although at small momentum transfer $|\vq| = |\vk - \vkp|$ (in the Laboratory frame), collective effects involving several nucleons should play a role. On the other side, as the excitation energy ($\omega=k_0-k_0'$) increases, inelastic channels 
\be
\nu_l (p) \,+ \,N(k) \raw  l^-(k')\, + \, X(p') 
\label{inel}
\ee
with $X=(\mathrm{ph}) \,N$, $\pi \,N$, $\dots$ start to open. As these processes are not always identified experimentally (pions can be absorbed and the nuclear products may not be all detected), they cannot be separated from CCQE in a model independent way. 

The MiniBooNE experiment, running with $\langle E_\nu \rangle \sim 750$~MeV on a CH$_2$ target, 
has collected the largest sample available so far for low energy $\nu_\mu$ CCQE~\cite{:2007ru}. After subtracting the non CCQE background, mainly from $\Delta(1232)$ excitation, using the NUANCE event generator~\cite{Casper:2002sd}, the CCQE data set was analyzed with the relativistic Global Fermi Gas model of Smith and Moniz (SM)~\cite{Smith:1972xh}. The shape of the muon angular and energy distributions averaged over the $\nu_\mu$ flux $\langle d\sigma / d\cos{\theta_\mu} dE_\mu \rangle$ could be described with rather standard values of the Fermi momentum $p_F = 220$~MeV and binding energy $E_B = 34$~MeV, but restricting the available phase space for the final proton by means of an ad hoc parameter $\kappa = 1.019 \pm 0.011$ such that $p'^0_{\mathrm{min}} = \kappa ( \sqrt{M^2 + p_F^2} - \omega + E_B )$, 
and taking $M_A = 1.23 \pm 0.20$~GeV~\cite{:2007ru}; this value of $M_A$ is in agreement with the K2K result $M_A = 1.2 \pm 0.12$~GeV~\cite{Gran:2006jn} on $^{16}$O, but considerably higher than the one obtained from $\nu_\mu$-deuterium data $M_A=1.0137 \pm 0.0264$~GeV~\cite{Bodek:2007vi} or by NOMAD at high energies (3-100~GeV) also on $^{12}$C [$M_A = 1.05 \pm 0.02(stat) \pm 0.06(syst)$~GeV]~\cite{Lyubushkin:2008pe}. A recent reanalysis using charged current single pion production (CC1$\pi$) data to adjust the Monte Carlo simulation employed to subtract the background obtains $\kappa = 1.007 \pm 0.007$ and $M_A = 1.35 \pm 0.17$~GeV~\cite{Katori:2009du}. While the first shape-only fit falls short compared to the measured integrated cross section~\cite{:2007ru}, the second one underestimates it only by 10~\%~\cite{Katori:2009du}.     

While such a modified SM model might be convenient to parametrize the CCQE cross section using a small number of parameters, it is important to understand the MiniBooNE CCQE data with more realistic nuclear models that implement the knowledge gathered through many years of research in electron-nucleus scattering. Besides, the fact that the description changes with the background subtraction procedure indicates the need of making theoretical predictions that could be compared to (more) inclusive and less model dependent data.

\section{The model}

The scattering amplitude for the elementary process [Eq.~(\ref{qe})] is proportional to the product of the leptonic and hadronic currents
\be
\mathcal{M} =  \frac{G_F \cos{\theta_C}}{\sqrt{2}}  l^\alpha  J_\alpha \,.
\label{ampl}
\ee
While the charged-current leptonic current is given by the Standard Model, the hadronic one can be written in terms of form factors that contain the information about nucleon properties 
\be
J_\alpha = \bar{u}(p')  
\left[ \left( \gamma_{\alpha} - \frac{q \hspace{-1.5mm}/ \,q_{\alpha}}{q^2}  \right)  F^V_1 + \frac{i}{2m_N} \sigma_{\alpha\beta} q^{\beta} F^V_2 -\gamma_{\alpha}\gamma_5 F_A - \frac{q_{\alpha}}{m_N} \gamma_5 F_P \right] u(p) \,.
\label{curr}
\ee
The vector form factors $F^V_{1,2}$ are obtained from electron scattering~\cite{Bodek:2007vi}; using PCAC $F_P$ can be expressed as a function of $F_A$, given by Eq.~(\ref{FA}) with $g_A=1.267$; for $M_A$ we adopt a value of 1~GeV, consistent with the world data. 

Our description of the CCQE reaction on nuclei is based on a Local Fermi Gas model, i.e. at each space point the initial-nucleon momentum distribution is given by a Fermi sphere $f(\vr,\vp)=\Theta (p_F (r) -|\vp|)$ with radius $p_F (r) = [\frac{3}{2} \pi^2 \rho(r)]^{1/3}$, where $\rho(r)$ is the empirical nuclear density.  A Pauli blocking factor for the final nucleon $P_{\mathrm {Pauli}} = 1 - \Theta (p_F (r) -|\vp|)$ also applies. 
Such a simple model already incorporates a space-momentum correlation which is absent for the Global Fermi Gas~\cite{Leitner:2008ue}, and provides a framework where more elaborated many body dynamics can be naturally incorporated. In contrast to the constant binding of the SM model, here all the nucleons, initial and final, are embedded in a density and momentum dependent potential $V(\vp,\vr)$ whose parameters have been fixed by proton-nucleus scattering data~\cite{Leitner:2008ue,Teis:1996kx}. As a consequence, the nucleons acquire effective masses $m_{\mathrm{eff}}(\vp,\vr)$ given by $\sqrt{\vp^2 + m_N^2} + V(\vp,\vr) = \sqrt{\vp^2 + m^2_{\mathrm{eff}}}$. 

The presence of nucleon-nucleon (NN) interactions inside nuclei implies that nucleon propagators are dressed with complex selfenergies $\Sigma$. This leads to spectral functions 
\be
S(p) = - \frac{1}{\pi} \frac{\mathrm{Im}  \Sigma(p)}{[p^2-m_N^2-\mathrm{Re} \Sigma(p)]^2 + [\mathrm{Im}  \Sigma(p)]^2} \,.
\label{SF}
\ee
As most of the nucleons in the nucleus can be described as occupying single-particle states in a mean field potential~\cite{Ankowski:2007uy} we can neglect NN interactions for the initial nucleons (holes) and take $\mathrm{Im}\Sigma =0$. Then, $S_h(p) \rightarrow \delta (p^2 - m_{\mathrm{eff}}^2)$ and we recover the description of the initial state outlined in the previous paragraph. On the contrary, for the final nucleons (particle states) NN interactions should be considered. For this purpose in Eq.~(\ref{SF}) we take $\mathrm{Im} \Sigma = - \sqrt{(p^2)} \Gamma_{\mathrm{coll}} (p,r)$, with the collisional broadening $\Gamma_{\mathrm{coll}} = \rho \sigma_{NN} v_{\mathrm{rel}}$ fixed according to the parametrizations of the Giessen Boltzmann-Uehling-Uhlenbeck (GiBUU) framework~\cite{GiBUU}. As for $\mathrm{Re} \Sigma$, it is obtained from $\mathrm{Im}\Sigma$ with a once-subtracted dispersion relation demanding that at the pole $p_0^{\mathrm{(pole)}} = \sqrt{\vp^2 + m_{\mathrm{eff}}^2}$. More details can be found in Ref.~\cite{Leitner:2008ue}. 
  
\begin{figure}[ht]
	\label{fig1}
  \includegraphics[width=.9\textwidth]{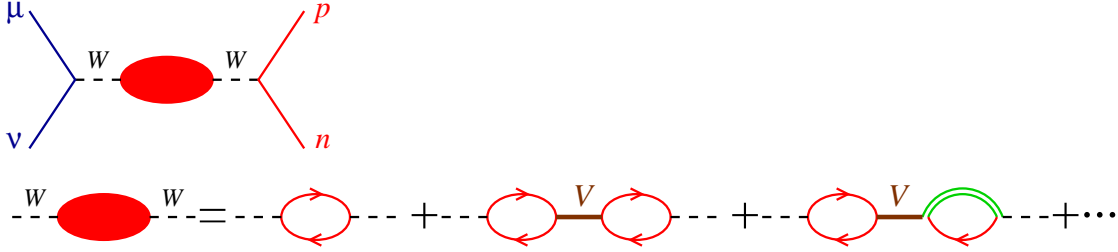}
  \caption{Long range correlations in CCQE scattering. Solid lines pointing to the right (left) denote particle (hole) states. The double line stands for the $\Delta(1232)$.}
\end{figure}

To complete our model  we take into account that inside nuclei, the strength of the electroweak couplings may change from their free nucleon values due to the presence of strongly interacting nucleons~\cite{Singh:1992dc}. In the nuclear medium, the axial coupling $g_A$ is renormalized in the same way as the electric field of a dipole is screened in a dielectric medium~\cite{EricsonWeise}  
\be
\frac{(g_A)_{\mathrm{eff}}}{g_A} = \frac{1}{1+g' \chi_0} \,,
\ee
where $\chi_0$ is the elementary dipole susceptibility and $g'$ the Lorentz-Lorenz factor whose classical value is $g' \sim 1/3$. This quenching of $g_A$ in nuclear Gamow-Teller $\beta$ decay is well established experimentally: $(g_A)_{\mathrm{eff}} / g_A \sim 0.9$~\cite{Wilkinson:1973} and was first applied to CCQE scattering by Singh and Oset~\cite{Singh:1992dc}. Such medium polarization effects involve several nucleons in the nucleus and are therefore important at low $|\vq|$ where the space resolution of the probe is large compared with the average NN separation. This corresponds to the region where MiniBooNE data exhibit a reduction with respect to the SM model handled in the analysis by introducing the $\kappa$ parameter. Following Nieves {\it et al.}~\cite{Nieves:2004wx}, we modify the lepton-nucleon interaction by an infinite sum of particle-hole (ph) states (RPA), as illustrated in Fig.~\ref{fig1}, 
which interact with an effective potential cast as \footnote{Only the terms that contribute to CCQE are shown.}
\be
\label{pot}
V=\btau_1 \btau_2 \sigma_1^i \sigma_2^j [\hat{q}_{i}\hat{q}_{j}V_{L}(q)
+({\delta}_{ij}- \hat{q}_{i}\hat{q}_{j})V_{T}(q) ] + c_0 f' \btau_1 \btau_2 \,;
\ee   
$V_L$ ($V_T$) contain explicit $\pi$ ($\rho$) exchange
\be
\label{ex}
V_L = \frac{f_{NN\pi}^2}{m^2_\pi}\left
\{\left(\frac{\Lambda_\pi^2-m_\pi^2}{\Lambda_\pi^2-q^2 }\right)^2
\frac{\vq{\,^2}}{q^2-m_\pi^2} + g^{\prime} \right \} \,, \,\,
V_T = \frac{f_{NN\pi}^2}{m^2_\pi}\left
\{C_\rho \left(\frac{\Lambda_\rho^2-m_\rho^2}{\Lambda_\rho^2-q^2 }\right)^2
\frac{\vq{\,^2}}{q^2-m_\rho^2} + g^{\prime} \right \} 
\ee
and a short range part effectively included in the phenomenological constant $g'$ with values in the range $g'=0.6 \pm 0.1$. Details about couplings and cutoff parameters $\Lambda_{\pi, \rho}$ can be found in Ref.~\cite{Nieves:2004wx}. No meson exchange is directly associated with the scalar term in Eq.~(\ref{pot}), assumed to be density dependent
\be
f'=\frac{\rho(r)}{\rho(0)} f'^{(in)} + \left[ 1 - \frac{\rho(r)}{\rho(0)} \right] f'^{(ex)} \,
\ee
where the parameters $f'^{(in)}= 0.33$, $f'^{(ex)}=0.45$ (and $c_0=380$~MeV~fm$^3$) are tuned to describe collective nuclear excitations~\cite{Speth:1980kw}. The RPA sum also includes $\Delta$-hole excitations as shown in Fig.~\ref{fig1}. The ph-$\Delta$h and $\Delta$h-$\Delta$h interactions can be obtained by replacing $\bsigma$ ($\btau$) with the spin (isospin) $1/2 \raw 3/2$ transition operators $\mathbf{S}$ ($\mathbf{T}$) in Eq.~(\ref{pot}) and $f_{NN\pi}$ by $f_{\Delta N\pi}$ in Eq.~(\ref{ex}). Explicit expressions for the RPA corrections to the hadronic tensor are given in Appendix A of Ref.~\cite{Nieves:2004wx}.

This RPA approach, built up with the single-particle states of the Local Fermi Gas is simpler than other more sophisticated methods such as continuum RPA and applies only to inclusive processes, but it incorporates explicitly $\pi$, $\rho$ exchange and $\Delta$-hole states and can be naturally inserted in a unified framework to study different neutrino-induced reactions like inclusive quasielastic scattering, nucleon knockout and pion production~\cite{Leitner:2006ww}. Moreover, it had been successfully applied to photo- and electro-nuclear reactions~\cite{Carrasco:1989vq,Gil:1997bm} and allows to describe simultaneously inclusive muon capture on $^{12}$C and the low energy LSND CCQE measurements (with $\nu_e$ and $\nu_\mu$)~\cite{Nieves:2004wx}. 

Finally, in order to compare to model-independent data it is necessary to include the contributions from the processes in Eq.~(\ref{inel}) that look like CCQE events in the detector. The main source of such a background is pion production from $\Delta(1232)$ excitation ($\nu_\mu \, N \raw \mu^- \, \Delta$) followed by  absorption ($\Delta \, N \rightarrow  N \, N$). Pion final state interactions
 in the nuclear medium are treated with a semiclassical BUU model in coupled channels (GiBUU, see Refs.~\cite{Leitner:2009ec,Leitner:2008wx} for details). 

\section{Results}

\begin{figure}[h]
	\label{fig2}
  \begin{minipage}{.49\textwidth}
    \includegraphics[width=.98\textwidth]{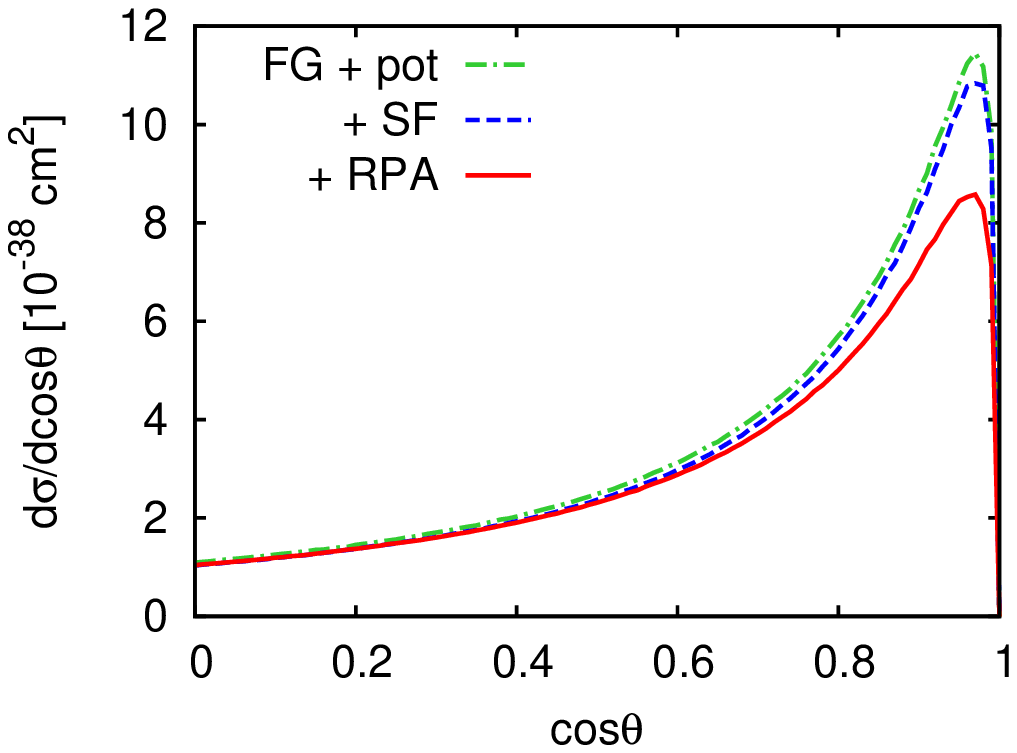}
    \includegraphics[width=.98\textwidth]{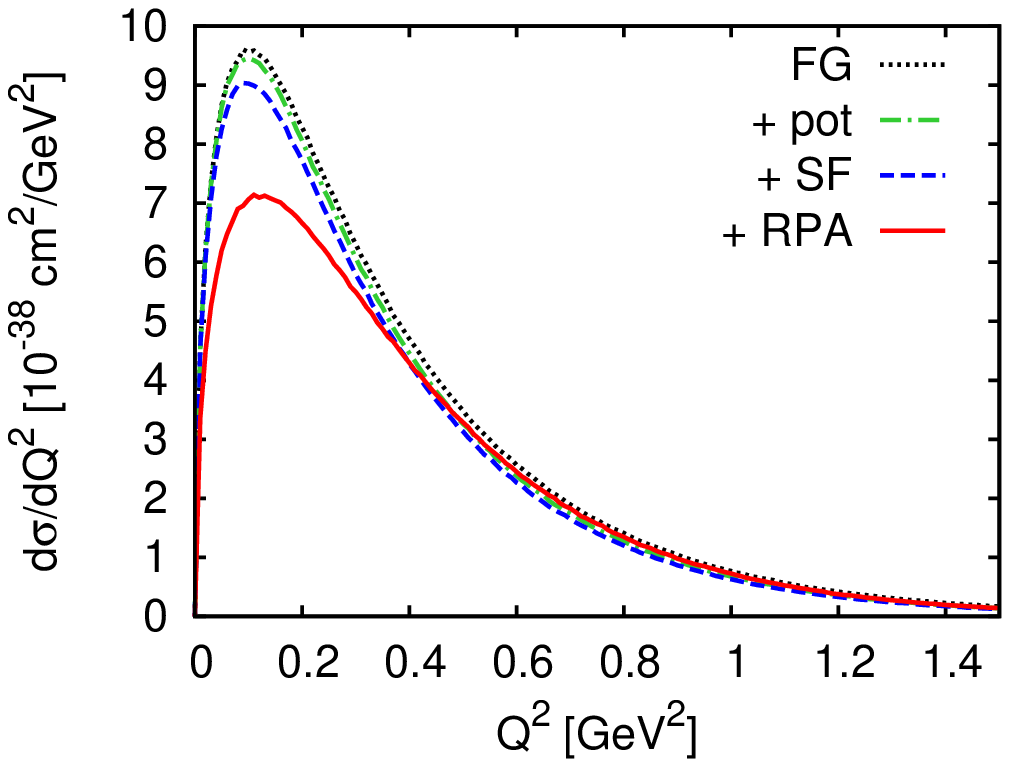}
  \end{minipage} \hfill	
  \begin{minipage}{.49\textwidth}
    \includegraphics[width=.98\textwidth]{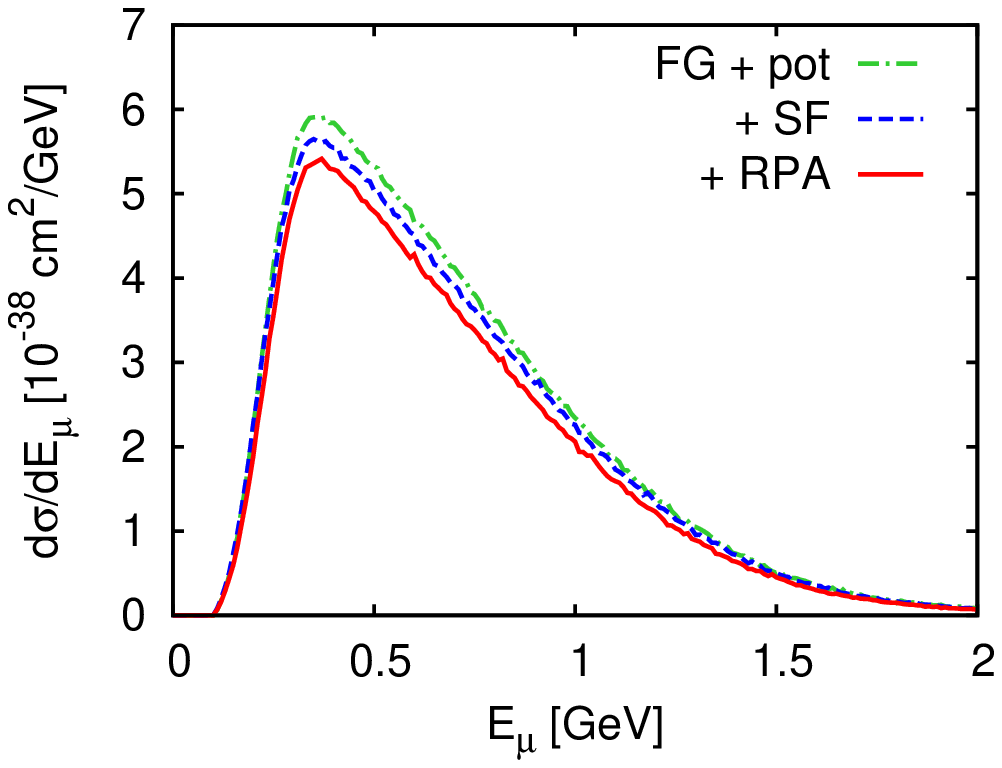}
    \includegraphics[width=.98\textwidth]{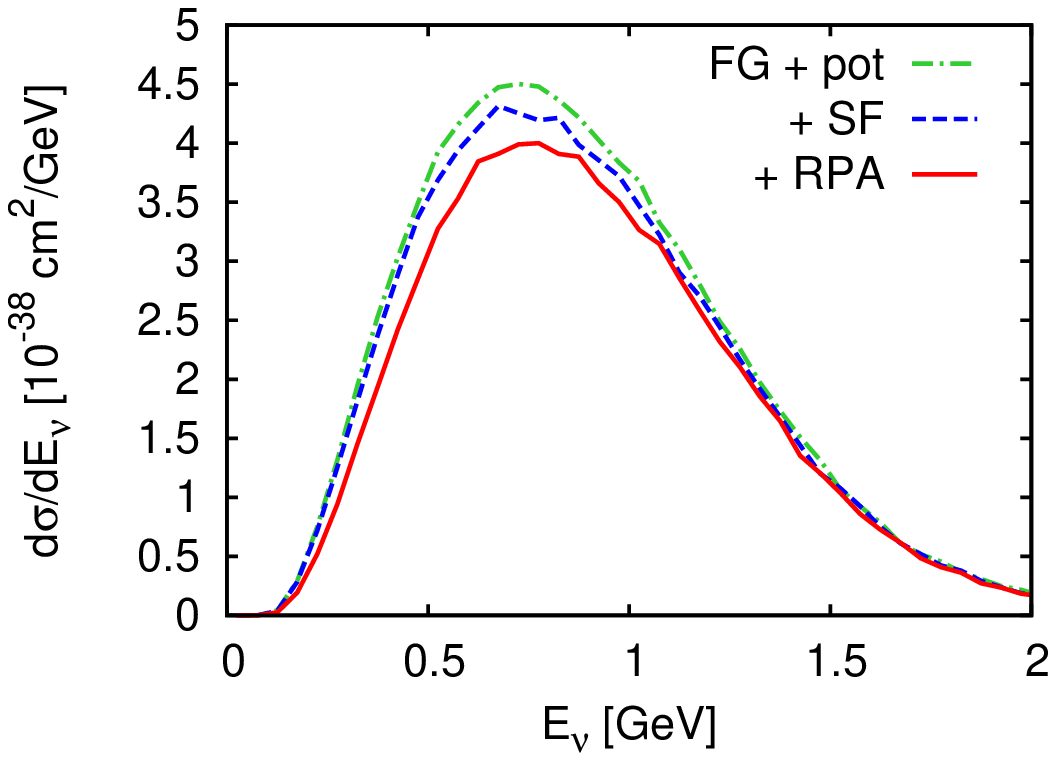}
  \end{minipage}
  \caption{Differential cross sections for the CCQE reaction~(\ref{qe}) on $^{12}$C and with $l=\mu$ averaged over the MiniBooNE flux as a function of the cosine of the outgoing muon angle (upper left), its energy (upper right), the (true) four-momentum transfer squared (lower left) and the (true) neutrino energy (lower right). Dotted lines represent the Local Fermi Gas model with Fermi motion and Pauli blocking. In the dash-dotted lines the nucleons are exposed to the mean field potential while the dashed ones also incorporate spectral functions for the outgoing nucleons. The full model with long range (RPA) correlations is denoted by solid lines.}
\end{figure}

In Fig.~\ref{fig2} we present the predictions of our CCQE model on $^{12}$C averaged over the MiniBooNE flux~\cite{AguilarArevalo:2008yp} for four different distributions. It is important to stress that, while the upper two correspond to directly measurable quantities (energy and scattering angle of the outgoing muons), the lower ones can only be experimentally obtained by reconstruction, assuming that the target nucleon is at rest~\cite{:2007ru}. Further corrections can be made by mapping reconstructed to true energy with the help of a reaction model, as done by MiniBooNE with their Monte Carlo simulation~\cite{AguilarArevalo:2009eb}.    
The plots reveal that the nuclear many-body corrections taken into account reduce the cross section. The effect of the spectral functions is rather small for these observables but the long range correlations cause a considerable reduction at forward angles ($\cos{\theta_{\mu}} < 0.8$) and low $Q^2 < 0.3$~GeV$^2$. 

\begin{figure}[h]
	\label{fig3}
    \includegraphics[width=.48\textwidth]{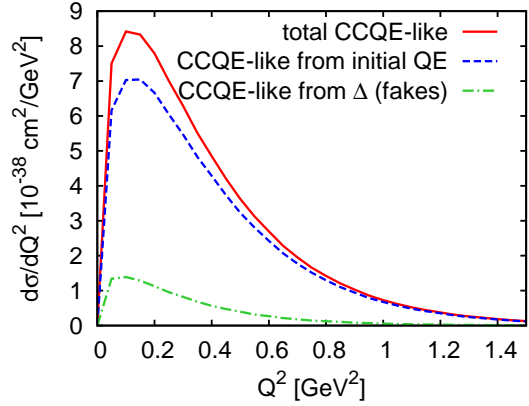}
  \caption{$Q^2$ distribution of CCQE-like events averaged over the MiniBooNE flux (solid line). It is given by the sum of the pure CCQE contribution (dashed line) plus fake CCQE events where the $\Delta$ is excited but no pion is produced (dash-dotted line)}.
\end{figure}

As discussed above, model-independent comparisons to data must include the fake CCQE background. Our prediction for this background from $\Delta$ excitation
and the full CCQE-like yield is shown in Fig.~\ref{fig3} for $d\sigma/dQ^2$. In total we obtain a fraction of fake CCQE over the total CCQE-like events of 10~\%, slightly smaller than the prediction of the MiniBooNE Monte Carlo: 12~\% (9.4~\% CC1$\pi^+$ resonant plus 2.5 ~\% CC1$\pi^0$)~\cite{AguilarArevalo:2009eb}. 

\begin{figure}[h]
	\label{fig4}
  \begin{minipage}{.49\textwidth}
    \includegraphics[width=.98\textwidth]{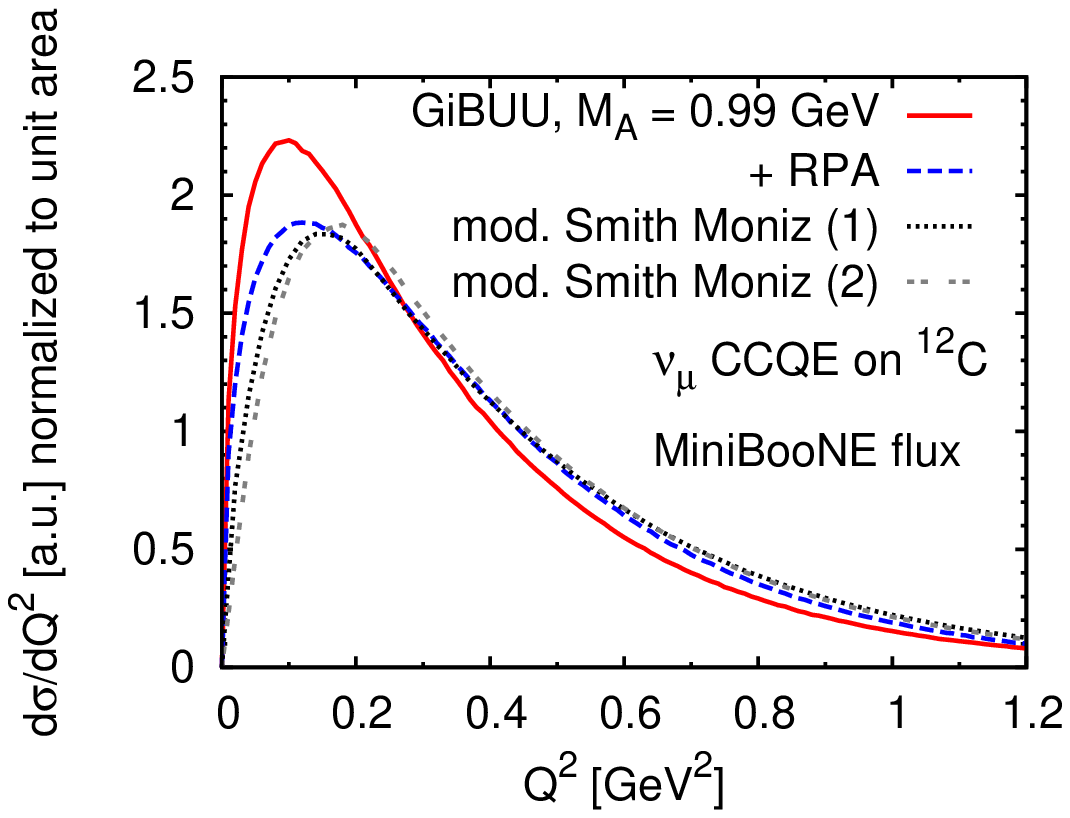}
 \end{minipage} \hfill	
  \begin{minipage}{.49\textwidth}
    \includegraphics[width=.98\textwidth]{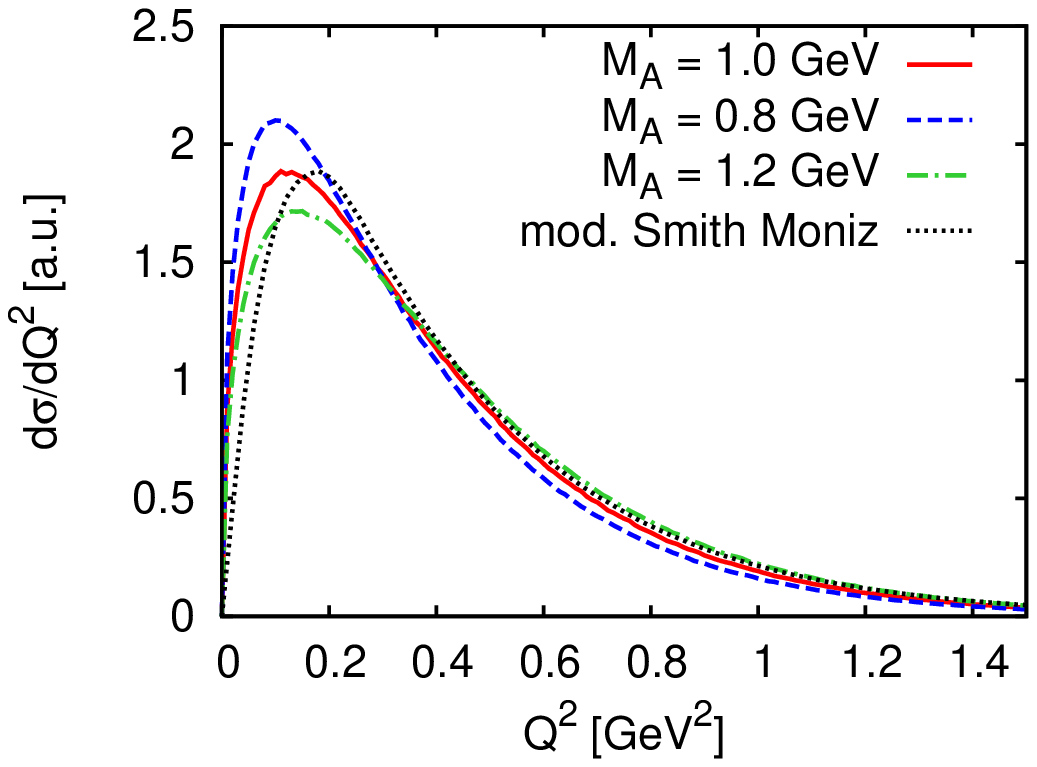}
 \end{minipage}
  \caption{Shape of the $Q^2$ distribution for the $\nu_\mu$-induced CCQE reaction on $^{12}$C averaged over the MiniBooNE flux. On the left panel, the prediction of our model without (solid line) and with RPA correlations (dashed line) is compared to the modified SM model with $\kappa = 1.007$, $M_A = 1.35$~GeV (1) and $\kappa = 1.019$, $M_A = 1.23$~GeV (2). On the right panel, the modified SM model (2) is confronted with the present model (including RPA) evaluated for different values of $M_A$. All curves are normalized to the same area.}  
  \end{figure}

CCQE-like data from MiniBooNE are not yet available. Nevertheless, it is useful to compare our pure CCQE predictions with the results of the MiniBooNE analysis that describe the shape of $\langle d\sigma / dQ^2 \rangle$ with a modified SM model~\cite{:2007ru,Katori:2009du}. Such a shape-only comparison for the $Q^2$ distribution is presented in the left panel of Fig.~\ref{fig4}. The two sets of ($\kappa, M_A$) values, the new $\kappa = 1.007$, $M_A = 1.35$~GeV [denoted as (1)] and the original $\kappa = 1.019$, $M_A = 1.23$~GeV  [denoted as (2)] are shown, both actually leading to very similar shapes. The comparison reveals that the RPA correlations, which reduce the size of the peak with respect to the tail,   
bring the shape of our distribution close to those extracted from data while keeping $M_A = 1$~GeV. 
On the other side, as it is clear from Fig.~\ref{fig2}, the integrated CCQE cross section obtained with our full model ($\langle \sigma \rangle = 3.2 \times 10^{-38}$~cm$^2$) is smaller than the one obtained within a standard Fermi gas model with $M_A = 1$~GeV. Instead, MiniBooNE gets a considerably larger value: $\langle \sigma \rangle = 5.65 \times 10^{-38}$~cm$^2$ with an error of 10.8~\%~\cite{Katori:2009du}. We have explored the sensitivity of the results to some of the uncertain magnitudes in the model. In particular, changing $g'$, whose contribution to the correlations is the largest at low $Q^2$, within acceptable values $g' = 0.5-0.7$ leaves the shape of the $Q^2$ distribution practically unchanged; the impact on the integrated cross section is also small: $\langle \sigma \rangle = 3.1-3.4 \times 10^{-38}$~cm$^2$. Increasing the value of $M_A$ causes an increase in the total cross section, but at the same time, the description of the shape gets worse, as illustrated on the right panel of Fig.~\ref{fig4}.

\section{Concluding remarks}

The theoretical model for the CCQE reaction presented here incorporates important many body corrections to the basic Local Fermi Gas picture like spectral functions and long range RPA correlations. The latter are important at low $Q^2$ where collective effects play a role. We find that a good agreement with the shape of the CCQE $Q^2$ distribution extracted from MiniBooNE data with $M_A = 1$~GeV, which is favored by early neutrino data, by the analysis of pion electroproduction close to threshold and by a recent neutrino experiment at high energies (NOMAD). However our description clearly underestimates the MiniBooNE integrated CCQE cross section. The situation is common to other models that take into account short-range correlations~\cite{Meloni} or apply the phenomenological scaling function extracted from electron scattering~\cite{Amaro:2006tf}. Other many-body mechanisms like meson exchange currents might add some additional strength and should be investigated. One should also bear in mind that the MiniBooNE result is not at all model independent. It relies on simulations to determine the neutrino beam and to subtract the fake contributions to the CCQE-like cross section. The plot of the MiniBooNE integrated CCQE cross section as function of the reconstructed neutrino energy together with NOMAD data (Fig.~6 of Ref.~\cite{Katori:2009du}) shows that, to make both experimental results compatible, the CCQE cross section would have to exhibit an unusual behavior, decreasing $20-30$~\% in the $E_\nu = 2-4$~GeV region to saturate afterwards.
Further joint theoretical and experimental work is necessary reconcile available CCQE theoretical calculations and experimental values.

\begin{theacknowledgments}
LAR thanks Juan Nieves for useful discussions and the S\'eneca Foundation for financial support during his stay in the University of Murcia. This work has been supported in part by the Deutsche Forschungsgemeinschaft. 
\end{theacknowledgments}

\end{document}